\documentclass[12pt]{article}
\usepackage{amsfonts}
\usepackage{amssymb,amsmath,amsthm,latexsym}
\usepackage[numbers,sort&compress]{natbib}
\usepackage{amsmath}
\usepackage{indentfirst}
\usepackage{adjustbox}

\usepackage[figuresright]{rotating}

\allowdisplaybreaks[4]
\newtheorem{thm}{Theorem}[section]
\newtheorem{prop}[thm]{Proposition}
\newtheorem{lem}[thm]{Lemma}
\newtheorem{cor}[thm]{Corollary}
\newtheorem{exam}{Example}
\newtheorem{defi}[thm]{Definition}

\newcommand{\pf}{{\bf Proof. \ }}

\numberwithin{equation}{section}
\begin{document}

\title{A new method for constructing EAQEC MDS codes}
\author{Peng Hu\\
 School of Mathematics and Physics, \\
 Hubei Polytechnic University  \\
 Huangshi, Hubei 435003, China, \\
 {Email: \tt HPhblg@126.com} \\
 Xiusheng Liu\footnote{Corresponding author.}\\
 School of Mathematics and Physics, \\
 Hubei Polytechnic University  \\
 Huangshi, Hubei 435003, China, \\
{Email: \tt lxs6682@163.com} \\}
\maketitle

%\vspace*{3cm}

\begin{abstract} Entanglement-assisted quantum error-correcting (EAQEC) codes make use of preexisting entanglement
between the sender and receiver to boost the rate of transmission. It is possible to construct an EAQEC code from
any classical linear code, unlike standard quantum error-correcting codes, which can only be constructed from dual-containing codes. However, the number $c$ of pre-shared maximally entangled states  is usually calculated by computer search. In this paper, we first give a new formula for calculating the number $c$ of pre-shared maximally entangled states. Then, using this formula, we construct three classes of new  entanglement-assisted quantum error-correcting maximum-distance-separable ( EAQEC MDS) codes.
\end{abstract}

%\vspace*{1cm}

\bf Key Words\rm :  Entanglement-assisted quantum MDS code $\cdot$ Rank of  matrix$\cdot$ Parity-check matrix

\bf Mathematics Subject Classification\rm :  94B05 $\cdot$ 94B25

\section{Introduction}
Quantum  codes are useful tool in quantum computation and communication to detect and correct the quantum errors while quantum information is transferred via quantum channel.  After the pioneering  work in \cite{AK01}, \cite{Calder1}, the theory of quantum  codes has developed rapidly in recent years. As we know, the approach of constructing new quantum codes which have good parameters is an interesting research field.   Many quantum codes  with good parameters were obtained from dual-containing classical linear codes concerning Euclidean inner product or Hermitian inner product (see \cite{AK01,AKS07,CLZ15,JLLX10,KZ1,Ket,LiuY}).

The previously mentioned  dual-containing conditions  prevent the usage of many common
classical codes for providing quantum codes. Brun et al. \cite{Brun01} proposed to share entanglement between
encoder and decoder to simplify the theory of quantum error correction and increase the communication capacity.
With this new formalism, entanglement-assisted quantum stabilizer codes can be constructed from any classical linear code
giving rise to entanglement-assisted quantum error-correcting (EAQEC) codes. Fujiwara et al. \cite{Fuji} gave a general
method for constructing entanglement-assisted quantum low-density parity check  codes.  Fan, Chen and Xu \cite{FanJ}
provided a construction of entanglement-assisted quantum maximum distance separable (EAQEC MDS) codes with a small number $c$ of
pre-shared maximally entangled states. From constacyclic codes, Chen et al. and Lu et al. constructed new EAQEC MDS codes with
larger minimum distance and consumed $4$ entanglement bits in \cite{Chen1,Lu1}, respectively. Let $c = 5$ and $c = 9$, Mustafa and Emre improved the parameters of EAQEC MDS codes with length $n$ further in \cite{Mu}.  Recently, in \cite{Liu,Liu1}, we  construct new EAQEC codes by using $s$-Galois dual codes and parts of them are EAQEC MDS codes.

Inspired by these works, in this paper,  we first give a new formula for calculating the number $c$ of
pre-shared maximally entangled states.  Then, using this formula, we construct new EAQEC MDS codes.

The paper is organized as follows. In Sect.2, we recall some basic knowledge on  linear codes, $s$-Galois dual codes and EAQEC codes. In Sect.$3$,  we give  a formula for calculating  the  number $c$ of pre-shared maximally entangled states by using generator matrix of one code and parity check matrix of another code.  And, in Sect.4, using the formula for calculating the number $c$, we obtain three classes of new EAQEC MDS codes.  Finally, some comparisons of EAQEC MDS codes and conclusions are made.

\section{Preliminaries}
In this section, we recall some basic concepts and results about linear codes, $s$-Galois dual codes,  and entanglement-assisted quantum error-correcting codes, necessary for the development of this work. For more details, we refer to \cite{Mac, Liu1, Fan, Fuji, Brun01, Bowen01, Lai01,Lai02, Wilde}.

Throughout this paper, let $p$ be a prime number and $\mathbb{F} _{q}$ be the finite field with $q=p^{e}$ elements, where $e$ is a positive number.  Let $\mathbb{F}_{q}^{*}$ be the multiplicative group of units of $\mathbb{F}_{q}$.

For a positive integer $n$, let $\mathbb{F}_{q}^n=\{{\bf x}=(x_1,\cdots,x_n)\,|\,x_j\in \mathbb{F}_{q}\}$
which is an $n$ dimensional vector  space over $\mathbb{F}_{q}$.  A linear $[n,k]_{q}$ code $C$ over $\mathbb{F} _{q}$ is  an $k$-dimensional subspace of $\mathbb{F}_{q}^{n}$. The Hamming weight $w_H(\mathbf{c})$ of a codeword $\mathbf{c} \in C$ is the number of nonzero components of $\mathbf{c}$. The Hamming distance of two codewords $\mathbf{c}_1,\mathbf{c}_2 \in C$  is $d_H(\mathbf{c}_1,\mathbf{c}_2) = w_H(\mathbf{c}_2-\mathbf{c}_1)$. The minimum Hamming distance  of $C$ is $d=\mathrm{min}\{w_H(\mathbf{a}-\mathbf{b})|\mathbf{a},\mathbf{b}\in C\}$. An $[n,k,d]_{q}$ code is an $[n,k]_{q}$ code with the minimum Hamming distance $d$. A $k\times n$ matrix $G$ over $\mathbb{F}_{q}$ is called a generator matrix of $C$, if the rows of $G$ generates $C$ and no proper subset of the rows of $G$ generates $C$.

\subsection{$s$-Galois dual codes}
Let  $s$  be integers with $0\leq s< e $ . In \cite{Fan}, Fan and Zhang introduced the following form
$$
[{\bf x},{\bf y}]_{s}=x_1y_1^{p^{s}}+\cdots+x_ny_n^{p^{s}},
\qquad\forall~ {\bf x},{\bf y}\in\mathbb{F}_{q}^n,
$$
where $q=p^e$ and $n$ is a positive integer.
We call $[{\bf x},{\bf y}]_{s}$ the $s$-Galois form on $\mathbb{F}_{q}^n$.
It is just the usual Euclidean inner product if $s=0$.
And, it is  the Hermitian inner product when $e$ is even and $s=\frac{e}{2}$.
For any code $C$ over $\mathbb{F}_{q}$ of length $n$, let
$$
C^{\bot_{s}}=\big\{{\bf x}\in\mathbb{F}_{q} ^n\,\big|\,
 [{\bf c},{\bf x}]_{s}=0,\, \forall~{\bf c}\in C\big\},
$$
which is called the $s$-Galois dual code of $C$. It is easy to check that $C^{\bot_{s}}$ is linear.
Then $C^{\bot_{0}}$ (simply, $C^{\perp}$) is just the Euclidean dual code of $C$,
and $C^{\bot_{\frac{e}{2}}}$ (simply, $C^{\perp_{H}}$)
is just the Hermitian dual code of $C$.
In particular, if $C \subset C^{\bot_{s}}$, then $C$ is $s$-Galois self-orthogonal.
Furthermore, we call $C$ is $k$-Galois self-dual if $C= C^{\bot_{s}}$.

In fact, the $s$-Galois form is non-degenerate (\cite[Remark 4.2]{Fan}).
This implies that
$\dim_{\mathbb{F}_{q}}C+\dim_{\mathbb{F}_{q}}C^{\perp_{s}}=n$.

For an $l\times n$ matrix $A =(a_{ij})_{l\times n}$ over $\mathbb{F}_{q}$, where $a_{ij}\in \mathbb{F}_{q}$,
we denote $A^{(p^{e-s})} = (a_{ij}^{p^{e-s}})_{l\times n}$, and $A^{T}$ as the transpose matrix of $A$.
Then for vetor $\mathbf{a}=(a_1,a_2,\ldots,a_n)\in\mathbb{F}_{q} ^n$,
we have
$$\mathbf{a}^{p^{(e-s)}}=(a_1^{p^{e-s}},a_2^{p^{e-s}},\ldots,a_n^{p^{e-s}}).$$

For a linear code $C$ of $\mathbb{F} _{q}^n$, we define $C^{(p^{e-s})}$
to be the set $\{\mathbf{a}^{p^{(e-s)}}\mid~\mathbf{a}\in C\}$ which is also a
linear code. It is easy to see that the $s$-Galois dual $C^{\perp_s}$ of $C$ is equal to the Euclidean
dual $(C^{(p^{e-s})})^{\perp}$ of the linear code $C^{(p^{e-s})}$.

%The following corollary is obvious.
%\begin{cor}\label{cor:2.1}
%If $C$ is an $[n,k,d]$ linear code  over $\mathbb{F}_{q}$ with a generating matrix $G$,
%enerating matrix $G^{(p^{e-k})}$. Moreover, $C$ is MDS if and only if $C^{\perp}$ is MDS.
%\end{cor}

\subsection{Entanglement-assisted quantum error-correcting codes}
An $[[n,k,d;c]]_q$ EAQEC code over $\mathbb{F}_{q}$ encodes $k$ logical qubits into $n$ physical qubits with the help of $c$ copies of maximally entangled states ($c$ ebits). The performance of an EAQEC code is measured by its rate $\frac{k}{n}$ and net rate $\frac{k-c}{n}$.

If $c=0$, then the EAQEC code is a standard stabilizer code.  EAQEC codes can be regarded as generalized quantum codes.

It has been prove that EAQEC codes have some advantages over standard stabilizer codes. In \cite{Wilde}, Wilde and Brun proved that EAQEC codes can be constructed using classical linear codes as follows.

\begin{prop} \label{prop:2.1} (\cite{Wilde})
Let $H_{1}$  and $H_2$ be parity check matrices of two linear codes $[n,k_1,d_1]_q$ and $[n,k_2,d_2]_q$,  respectively. Then an $[[n,k_1+k_2-n+c,\mathrm{min}\{d_1,d_2\};c]]_q$ EAQEC code can be obtained, where $c=\mathrm{rank}(H_1H_2^{T})$ is the required number of maximally entangled states.
\end{prop}

To see how  good an EAQEC code is in terms of its parameters, we use the entanglement-assisted quantum Singleton bound.
\begin{thm} \label{theorem:2.1}(\cite{Brun01})
For any $[[n,k,d;c]]_{q}$ EAQEC code with $0\leq c\leq n-1$, it holds that $2(d-1)\leq n-k+c$.
\end{thm}

If  an  EAQEC code $Q$ with parameters $[[n,k,d;c]]_{q}$ attains the entanglement-assisted quantum Singleton bound $2(d-1)= n-k+c$, then it is called the entanglement-assisted quantum  maximum-distance-separable (EAQEC MDS) code.

%\begin{defi}\label {de:2.2}
%Let $Q$ be an EAQECC with parameters $[[n,k,d;c]]_{q}$. If  $c=n-k$, it is called a maximal-entanglement EAQECC.
%\end{defi}

\section{A new formula for calculating the number $c$}

 We first verify the following a lemma.
\begin{lem}\label{le:3.1}
Let $C$ be an $[n,k]_{q}$ linear code over $\mathbb{F}_{q}$ with generator matrix $G$ and parity check matrix $H$.  Then $$(C^{(p^{e-s})})^{\perp}=(C^{\perp})^{(p^{e-s})}.$$
\end{lem}
\pf By assumptions, it is easy to prove that the matrix $G^{(p^{e-s})}$ is a generator matrix of the linear code $C^{(p^{e-s})}$, and the matrix $H^{(p^{e-s})}$ is a generator matrix of the linear code $(C^{\perp})^{(p^{e-s})}$.

Let $\mathbf{g}_1,\ldots,\mathbf{g}_k$ be rows of the $G$, and let $\mathbf{h}_1,\ldots,\mathbf{h}_{n-k}$  be rows of the $H$.
For any $\mathbf{x}\in(C^{\perp})^{(p^{e-s})}$, we can assume that
$$\mathbf{x}=y_1\mathbf{h}_1^{p^{e-s}}+\cdots+h_{n-k}\mathbf{h}_{n-k}^{p^{e-s}}.$$
Then, for any $\mathbf{g}_j^{p^{e-s}}\in G^{(p^{e-s})}$, we have
$$[\mathbf{x},\mathbf{g}_j^{p^{e-s}}]=\sum_{i=1}^{n-k}y_i[\mathbf{h}_i^{p^{e-s}},\mathbf{g}_j^{p^{e-s}}]=\sum_{i=1}^{n-k}y_i[\mathbf{h}_i,\mathbf{g}_j]^{p^{e-s}}=0.$$
Therefore, $\mathbf{x}\in(C^{(p^{e-s})})^{\perp}$, which implies
$$(C^{\perp})^{(p^{e-s})}\subset(C^{(p^{e-s})})^{\perp}.$$

Clearly, $\mathrm{dim}_{\mathbb{F}_q}(C^{\perp})^{(p^{e-s})}=\mathrm{dim}_{\mathbb{F}_q}(C^{(p^{e-s})})^{\perp}$.

Summarize,
$$(C^{(p^{e-s})})^{\perp}=(C^{\perp})^{(p^{e-s})}.$$
\qed
\begin{cor} \label{cor:3.1}
Let $C_{i}$   linear code $[n,k_i,d_i]_q$ over $\mathbb{F}_{q}$ with  parity check matrix $H_i$ for $i=1,2$. Then an $[[n,k_1+k_2-n+c,\mathrm{min}\{d_1,d_2\};c]]_q$ EAQEC code can be obtained, where $c=\mathrm{rank}(H_1(H_2^{(p^{e-s})})^{T})$ is the required number of maximally entangled states.
\end{cor}
\pf By Lemma \ref{le:3.1}, $H_2^{(p^{e-s})}$ is a parity check matrix of the code $C^{p^{e-s}}$. It is easy to prove that  code $C^{p^{e-s}}$ is a linear code with parameters $[[n,k_2,d_2]_q$. Then, in light of Proposition \ref{prop:2.1}, there exists an EAQEC code with parameters $[[n,k_1+k_2-n+c,\mathrm{min}\{d_1,d_2\};c]]_q$,  where $c=\mathrm{rank}(H_1(H_2^{(p^{e-s})})^{T})$ is the required number of maximally entangled states.
\qed

\begin{lem}\label{le:3.2}
Let $C_i$ be an $[n,k_i]_{q}$ linear code over $\mathbb{F}_{q}$ with generator matrix $G_i=\begin{pmatrix}\mathbf{g}_{i,1}\\\mathbf{g}_{i,2}\\ \vdots \\ \mathbf{g}_{i,k_i}\end{pmatrix}$ and parity check matrix $H_i=\begin{pmatrix}\mathbf{h}_{i,1}\\\mathbf{h}_{i,2}\\ \vdots \\ \mathbf{h}_{i,n-k_i}\end{pmatrix}$for $i=1,2$.  Then $$\mathrm{dim}_{F_q}(C_1\cap C_2^{\perp_s})=k_1+n-k_2-~\mathrm{rank}\begin{pmatrix}G_1\\H_2^{(p^{e-s})}\end{pmatrix}.$$
\end{lem}
\pf Let $\mathbf{a}\in C_1\cap C_2^{\perp_s}$. Then there exist $x_1,\ldots ,x_{k_1},y_1,\ldots ,y_{n-k_2}\in \mathbb{F}_{q}$ such that
$$x_1\mathbf{g}_{1,1}+\cdots+x_{k_1} \mathbf{g}_{1,k_1}=-y_1\mathbf{h}_{2,1}^{p^{e-s}}+\cdots-y_{n-k_2} \mathbf{h}_{2,n-k_2}^{p^{e-s}},$$
that is, $(x_1,\ldots ,x_{k_1},y_1,\ldots, y_{n-k_2})$ is the solution of a system of linear equations
$$x_1\mathbf{g}_{1,1}+\cdots+x_{k_1} \mathbf{g}_{1,k_1}+y_1\mathbf{h}_{2,1}^{p^{e-s}}+\cdots+y_{n-k_2}\mathbf{h}_{2,n-k_2}^{p^{e-s}}=0.$$
Thus, $\mathrm{dim}_{F_q}(C_1\cap C_2^{\perp_s})=k_1+n-k_2-
~\mathrm{rk}(G_1^T|(H_2^{(p^{e-s})}))^T=k_1+n-k_2-~\mathrm{rank}\begin{pmatrix}G_1\\H_2^{(p^{e-s})}\end{pmatrix}$.  This proves
the expected result.
\qed

In terms of the generator matrix of one linear code $C_1$ and  the parity-check
matrix of another linear code $C_2$ over $\mathbb{F}_q$, we now give a new formula for computing the number $c$ of pre-shared maximally entangled states.

\begin{thm}\label{th:3.1}
Let $C_i$ be an $[n,k_i]_{q}$ linear code over $\mathbb{F}_{q}$ with generator matrix $G_i$ and parity check matrix $H_i$ for $i=1,2$. Then $$c=\mathrm{rank}(H_1(H_2^{(p^{e-s})})^{T})=\mathrm{rank}\begin{pmatrix}G_1\\H_2^{(p^{e-s})}\end{pmatrix}-k_1.$$
\end{thm}
\pf By Lemma \ref{le:3.1}, we have $C_2^{\perp_s}=(C_2^{(p^{e-s})})^{\perp}=(C_2^{\perp})^{(p^{e-s})}$. Thus, $H_2^{(p^{e-k})}$ is a  generator matrix of $C_2^{\perp_s}$, i.e., $H_2^{(p^{e-k})}$ is a parity check matrix of $C_2^{(p^{e-s})}$.

Let $\mathbf{h}_{i,1},\mathbf{h}_{i,2},\ldots, \mathbf{h}_{i,n-k_i}$ be rows of the parity check matrix $H_i$ for $i=1,2$. Then $\mathbf{h}_{i,1}^{p^{e-s}}, \mathbf{h}_{i,2}^{p^{e-s}},\ldots, \mathbf{h}_{i,n-k_i}^{p^{e-s}}$ are  rows the parity check $H_i^{(p^{e-s})}$ for $i=1,2$.

Let $\sum_{j=1}^{n-k_2} x_j\mathbf{h}_{2,j}^{p^{e-s}}\in C_2^{\perp_s}$ , where $x_j\in\mathbb{F}_{q}$ for all $1\leq j\leq n-k_2$. Then $\sum_{j=1}^{n-k_2} x_j\mathbf{h}_{2,j}^{p^{e-s}}\in C_1\cap C_2^{\perp_s}$ if and only if for any $t\in \{1,2,\ldots,k_1\}$, we have
$$[\sum_{j=1}^{n-k_2} x_j\mathbf{h}_{2,j}^{p^{e-s}},\mathbf{h}_{1,t}]=0,$$
that is
$$\mathbf{x}H_2^{(p^{e-s})}H_1^{T}=0,$$
where $\mathbf{x}=(x_1,\ldots,x_{n-k_2})$. Therefore,
$$\mathrm{rank}(H_1(H_2^{(p^{e-s})})^{T})=\mathrm{rank}(H_2^{(p^{e-s})}H_1^{T})=n-k_2-\mathrm{dim}_{\mathbb{F}_q}(C_1\cap C_2^{\perp_s}).$$

In light of Lemma \ref{le:3.2}, we have
$$\mathrm{dim}_{\mathbb{F}_q}(C_1\cap C_2^{\perp_s})=n+k_1-k_2-\mathrm{rank}\begin{pmatrix}G_1\\H_2^{(p^{e-k})}\end{pmatrix}.$$
Thus,
$$c=\mathrm{rank}(H_1(H_2^{(p^{e-s})})^{T})=\mathrm{rank}\begin{pmatrix}G_1\\H_2^{(p^{e-k})}\end{pmatrix}-k_1.$$
\qed

\section{ Construction of EAQEC codes}
In this section, we give three classes of EAQEC MDS codes.

Combining Corollary \ref{cor:3.1} and Theorem \ref{th:3.1}, we can immediately get the following theorem.
\begin{thm} \label{th:4.1}
Let $G_{1}$  be a generator matrix of the $C_1=[n,k_1,d_1]_q$,  and let $H_2$ be a parity check matrix of the linear code  $C_2=[n,k_2,d_2]_q$. Then an $[[n,k_1+k_2-n+c,\mathrm{min}\{d_1,d_2\};c]]_q$ EAQEC code can be obtained, where $c=\mathrm{rank}\begin{pmatrix}G_1\\H_2^{(p^{e-k})}\end{pmatrix}-k_1$ is the required number of maximally entangled states.
\end{thm}

\subsection{The first classes of EAQEC MDS codes}

To construct a class of  new EAQEC MDS codes by using Theorem \ref {th:4.1}, we consider the Vandermonde matrices.

A Vandermonde $n\times n$ matrix $V_n=V(a_1,\ldots,a_n)$ is defined by
$$V_n=V(a_1,\ldots,a_n)=\begin{pmatrix}1&a_1&a_1^2&\cdots&a_1^{n-1}\\1&a_2&a_2^2&\cdots&a_2^{n-1}\\\vdots&\vdots&\vdots&\cdots&\vdots\\1&a_n&a_n^2&\cdots&a_n^{n-1}\end{pmatrix},$$
where $a_1,a_2,\ldots,a_{n}$ are elements of $\mathbb{F}_q^{*}$. It is well-known that the determinant of $V_n$ is non-zero if and only if the $a_i$ are distinct.

We recall the following fact (see \cite{Hur}).

\begin{lem} (\cite{Hur})\label{le:4.2} Let $C$ be a code generated by taking $k$ consecutive rows of a Vandermonde $n\times n$ matrix. Then $C$ is an MDS code with parameters $[n,k,n-k+ 1]_q$.
\end{lem}

\begin{thm} \label{th:4.3} Let $n\leq q-1$, $1\leq t\leq k+1$ and $k+1\leq t+j\leq n $. Then

$(1)$ there is an  EAQEC code with parameters $[[n,t-1,\mathrm{min}\{n-k+1,j+2\};j-k+t]]_q$.

$(2)$ when $n-k=1+j$, there is an  EAQEC MDS code with parameters $[[n,t-1,n-k+1;j-k+t]]_q$.
\end{thm}

\pf $(1)$ For $0<k<n$, take
$$G_{1}=\begin{pmatrix}1&a_1&a_1^2&\cdots&a_1^{n-1}\\1&a_2&a_2^2&\cdots&a_2^{n-1}\\\vdots&\vdots&\vdots&\cdots&\vdots\\1&a_k&a_k^2&\cdots&a_k^{n-1}\end{pmatrix}.$$
Let $C_1$ be a linear code with the generator matrix $G_{1}$. Then, by Lemma \ref{le:4.2}, $C_1$ is an MDS code with parameters $[n,k,n-k+ 1]_q$.

Take
$$H_{2}=\begin{pmatrix}1&a_t&a_t^2&\cdots&a_t^{n-1}\\1&a_{t+1}&a_{t+1}^2&\cdots&a_{t+1}^{n-1}\\\vdots&\vdots&\vdots&\cdots&\vdots\\1&a_{t+j}&a_{t+j}^2&\cdots&a_{t+j}^{n-1}\end{pmatrix}.$$
where $1\leq t\leq k+1$ and $k+1\leq t+j\leq n $.
Let $C_2$ be a linear code with the parity-check matrix $H_{2}$.  Then, again by Lemma \ref{le:4.2}, $C_2$ is an MDS code with parameters $[n,n-j-1,j+2]_q$.

Since $1\leq t\leq k+1$ and $k+1\leq t+j\leq n $, we have
$$c=\mathrm{rank}\begin{pmatrix}G_1\\H_2^{(p^{e-0})}\end{pmatrix}-k=j-k+t.$$
Thus, by Theorem \ref{th:4.1},  there exists an EAQEC code with parameters $[[n,t-1,\mathrm{min}\{n-k+1,j+2\};j-k+t]]_q$.

$(2)$ When $n-k=1+j$, according to $(1)$,  there is an EAQEC code with parameters $[[n,t-1,d;j-k+t]]_q$, where $d=\mathrm{min}\{n-k+1,j+2\}=n-k+1$.

Since $2(d-1)=2(n-k)=n-(t-1)+(j-k+t)$,  there is an EAQEC MDS code with parameters  $[[n,t-1,n-k+1;j-k+t]]_q$.
\qed

\begin{exam} By Theorem \ref{th:4.3}, taking some special $q$, we obtain  new
EAQEC MDS codes  in Table 1. Compared to the EAQEC MDS codes in \cite{Luo}, we have that the distance of our EAQEC MDS codes  obtained in Table 1 are larger than all of them.

\begin{table}
\caption{MDS EAQEC codes comparison.}
\begin{center}\begin{tabular}{|c|c|c|c|c|c|c|}
\hline
$q$&n&k&t&j&New  EAQEC MDS codes &EAQEC MDS codes \cite{Luo} \\
\hline
$13$&$12$&$4$&$5$&$7$&$[[12,4,9;8]]_{13}$&$[[12,4,7;4]]_{13}$\\&$12$&$5$&$6$&$6$&$[[12,5,8;7]]_{13}$&$[[12,5,7;5]]_{13}$
\\&$12$&$6$&$7$&$5$&$[[12,6,7;6]]_{13}$&$[[12,6,6;4]]_{13}$\\&$12$&$8$&$9$&$3$&$[[12,8,5;4]]_{13}$&$[[12,8,4;2]]_{13}$\\
\hline
$27$&$15$&$2$&$3$&$12$&$[[15,2,14;13]]_{27}$&$[[15,2,13;11]]_{27}$\\&$15$&$3$&$4$&$11$&$[[15,3,13;12]]_{27}$&$[[15,3,12;10]]_{27}$
\\&$15$&$4$&$5$&$10$&$[[15,4,12;11]]_{27}$&$[[15,4,11;9]]_{27}$\\&$15$&$5$&$6$&$9$&$[[15,5,11;10]]_{27}$&$[[15,5,10;8]]_{27}$
\\&$15$&$6$&$7$&$8$&$[[15,6,10;9]]_{27}$&$Not$\\&$15$&$7$&$8$&$7$&$[[15,7,9;8]]_{27}$&$[[15,7,7;4]]_{27}$\\&$15$&$8$&$9$&$6$&$[[15,8,8;7]]_{27}$&$[[15,8,7;5]]_{27}$
\\&$15$&$9$&$10$&$5$&$[[15,9,7;6]]_{27}$&$[[15,9,6;4]]_{27}$\\&$15$&$10$&$11$&$4$&$[[15,10,6;5]]_{27}$&$[[15,10,5;3]]_{27}$
\\&$15$&$11$&$12$&$3$&$[[15,11,5;4]]_{27}$&$[[15,11,4;2]]_{27}$\\
\hline
\end{tabular}\end{center}
\end{table}
\end{exam}

\subsection{The second classes of EAQEC MDS codes}

We now recall some basic results of Generalized Reed-Solomon  codes (see\cite{Jin2}). Let $\alpha_1,\ldots,\alpha_n$ be $n$ distinct elements of $\mathbb{F}_{q}$, and let  $v_1,\ldots,v_n$ be $n$ nonzero elements of $\mathbb{F}_{q}$. For $k$ between $1$ and $n$, the generalized Reed-Solomon code $GRS_k(\mathbf{\alpha},\mathbf{v})$ is defined by
$$GRS_k(\mathbf{a},\mathbf{v})=\{(v_1f(\alpha_1),\ldots,v_nf(\alpha_n))|~f(x)\in\mathbb{F}_{q}[x],deg(f(x))\leq k-1\},$$
where $\mathbf{a}$, $\mathbf{v}$ denote the vectors $(\alpha_1,\ldots,\alpha_n)$, $(v_1,\ldots,v_n)$, respectively.

Furthermore we consider the extended code of the generalized Reed-Solomon code $GRS_k(\mathbf{a}, \mathbf{v})$ given by
$$GRS_k(\mathbf{a},\mathbf{v},\infty)= \{(v_f(\alpha_1), v_2f(\alpha_2),\ldots , v_nf(\alpha_n), f_{k-1}) : f(x) \in\mathbb{F}_q[x], \mathrm{deg}(f(x)) \leq k-1\}.$$
where $f_{k-1}$ stands for the coefficient of $x^{k-1}$. The following two results can be found in \cite{Jin2}.
\begin{lem} (\cite{Jin2}) The code $GRS_k(\mathbf{a},\mathbf{v},\infty)$  is a MDS code with parameters $[n + 1, k, n-k+2]_q$.
\end{lem}

\begin{lem}(\cite{Jin2}) \label{le:4.5}Let $\mathbf{1}$  be all-one word of length $n$.   If $1 \leq k \leq q-1$, then the dual code of $GRS_k(\mathbf{a},\mathbf{v},\infty)$ is $GRS_{q-k+1}(\mathbf{a},\mathbf{v},\infty)$.
\end{lem}
\begin{thm}\label{th:4.6} Let  $1\leq k< \lceil\frac{q+1}{2}\rceil$. Then
there is an EAQEC MDS code with parameters $[[q+1,1,q-k+2;q-2k+2]]_q$.
\end{thm}
\pf Taking $$G_1=\begin{pmatrix}1&1&\cdots&1&0&\\\alpha_1&\alpha_2&\cdots&\alpha_q&0\\\alpha_1^2&\alpha_2^2&\cdots&\alpha_q^2&0\\ \vdots& \vdots&\ddots &\vdots &\vdots\\\alpha_1^{k-1}&v_2\alpha_2^{k-1}&\cdots&\alpha_q^{k-1}&1\end{pmatrix}.$$
Then, $G_1$ is a generator matrix of  $GRS_k(\mathbf{a},\mathbf{v},\infty)$.

Set, $$H_2=\begin{pmatrix}1&1&\cdots&1&0&\\\alpha_1&\alpha_2&\cdots&\alpha_q&0\\\alpha_1^2&\alpha_2^2&\cdots&\alpha_q^2&0\\ \vdots& \vdots&\ddots &\vdots &\vdots\\\alpha_1^{q-k}&v_2\alpha_2^{q-k}&\cdots&\alpha_q^{q-k}&1\end{pmatrix}.$$
Then, by Lemma \ref{le:4.5}, $H_2$ is a parity-check matrix of  $GRS_k(\mathbf{a},\mathbf{v},\infty)$.

Since $1\leq k< \lceil\frac{q+1}{2}\rceil$, we have
$$c=\mathrm{rank}\begin{pmatrix}G_1\\H_2^{(p^{e-0})}\end{pmatrix}-k=q-2k+2.$$
Thus, by Theorem \ref{th:4.1}, there exists an EAQEC code with parameters $[[q+1,1,q-k+2;q-2k+2]]_q$.

Since $2(d-1)=2(q-k+1)=q+1-1+(q-2k+2)$, the EAQEC code with parameters $[[q+1,1,q-k+2;q-2k+2]]_q$ is an EAQEC MDS code.
\qed
\begin{exam} By Theorem \ref{th:4.6}, taking some special $q$, we obtain  new
EAQEC MDS codes  whose parameters are $[[10,1,7;3]]_{9}$,$[[12,1,10;7]]_{11}$, $[[14,1,9;3]]_{13}$,$[[18,1,8;1]]_{17}$.
\end{exam}

\subsection{The third classes of EAQEC MDS codes}
In subsection, we assume that $q=l^m$ with $l$ prime power.

For brevity, we will use notion $[i]=l^{i~\mathrm{mod}~m}$,  $a^{[i]}=a^{l^{i~\mathrm{mod}~m}}$ for $a\in\mathbb{F}_{q}$ and integer $i$, where mod operation returns non negative value.

Given a vector  $(g_1,g_2,\ldots,g_n)\in \mathbb{F}_{q}^n$,  we denote by $M_k(g_1,g_2,\ldots,g_n)\in\mathbb{F}_{q}^{k\times n}$ the matrix
$$~~~~~~~~~~~~~~M_k(g_1,g_2,\ldots,g_n)=\begin{pmatrix}g_1 & g_2 & \ldots &g_n \\
     g_1^{[1]} & g_2^{[1]} & \ldots &g_n^{[1]}\\
     g_1^{[2]} & g_2^{[2]} & \ldots &g_n^{[2]}\\
    \vdots& \vdots & \ddots & \vdots \\
   g_{1}^{[(k-1)]} & g_2^{[(k-1)]} & \ldots &g_n^{[(k-1)]}\\
  \end{pmatrix}.~~~~~~~~~~~~~~~~~~~~~~~$$

A  definition of rank-metric code, proposed by Gabidulin, is the following.

\begin{defi} \label{de: 2} {\rm(\cite{Gab})} The rank of a vector $\mathbf{g}=(g_1,g_2,\ldots,g_n),g_i\in \mathbb{F}_{q}$, denoted by $\mathrm{rank}(\mathbf{g})$, is defined as the maximal number of linearly independent coordinates $g_i$ over $\mathbb{F}_{q}$, i.e., $\mathrm{rank}(\mathbf{g}):=\mathrm{dim}_{\mathbb{F}_q}\langle g_1,g_2,\ldots,g_n\rangle$. Then we have a metric rank  distance given by $d_{\mathrm{r}}(\mathbf{a}-\mathbf{b})=\mathrm{rk}(\mathbf{a}-\mathbf{b})$ for $\mathbf{a},\mathbf{b}\in\mathbb{F}_{q}^n $.  A Gabidulin (rank-metric) code of length $n$ with dimension $k$ over $\mathbb{F}_{q}$ is an $\mathbb{F}_{q}$-linear subspace $C\subset\mathbb{F}_{q}^n$. The minimum rank distance of a Gabidulin  code $C\neq0$ is
$$d_{\mathrm{r}}(C):=\mathrm{min}\{\mathrm{rank}(\mathbf{a}):~\mathbf{a}\in C,~\mathbf{a}\neq0\}.$$
\end{defi}

The Singleton bound for codes in the Hamming metric  implies also an upper bound  for Gabidulin codes.

\begin{thm} \label{th:Gab1} {\rm(\cite{Gab})}
Let $C\subset\mathbb{F}_{q}^n$ be a Gabidulin  code with minimum rank distance $d_r(C)$ of dimension $k$.  Then $d_r(C)\leq n-k+1.$
\end{thm}

A  Gabidulin  code attaining the Singleton bound  is called a Gabidulin maximum rank distance  (MRD) code.

In  paper \cite{KGab}, Kshevetskiy and Gabidulin  showed the following result on MRD codes:
\begin{thm}\label{th:Gab2}
Let $g_1,g_2,\ldots,g_n\in\mathbb{F}_{q}$ be linearly independent over $\mathbb{F}_{l}$, and let $C$ be a Gabidulin  code generated by matrix $M_k(g_1,g_2,\ldots,g_n)$. Then Gabidulin  code $C$ is an MRD code with parameters $[n,k,n-k+1]$.
\end{thm}

When $n\leq m$, $d_r(C)\leq d(C)$, where  $d(C)$ is the
minimum Hamming distance of $C$. Therefore, we have the following corollary.
\begin{cor} \label{co:Gab} Let $n\leq m$. If  $C$ is an MRD code with parameters $[n,k,n-k+1]$ over $\mathbb{F}_{q}$, then $C$ is also an MDS code with parameters $[n,k,n-k+1]$ over $\mathbb{F}_{q}$.
\end{cor}

\begin{cor}\label{co:Gab1}Let $n\leq m$. Let $g_1,g_2,\ldots,g_n\in\mathbb{F}_{q}$ be linearly independent over $\mathbb{F}_{l}$, and let $C$ be a Gabidulin  code generated by matrix $M_k(g_1^{l^t},g_2^{l^t},\ldots,g_n^{l^t})$, where $1\leq t\leq m-1$. Then Gabidulin  code $C$ is an MRD code with parameters $[n,k,n-k+1]$. Furthermore,  $C$ is also an MDS code with parameters $[n,k,n-k+1]$ over $\mathbb{F}_{q}$.
\end{cor}
\pf  We first verify that if $g_{1},g_{2}, \ldots, g_{n}$ are linearly independent over $\mathbb{F}_{l}$ then  $g_{1}^{l^t},g_{2}^{l^t}, \ldots, g_{n}^{l^t}$ are also linearly independent over $\mathbb{F}_{l}$. We prove it by contradiction. Suppose that there is not all zero  $a_{1},a_2,\cdots,a_n\in\mathbb{F}_{l}$ such that
$$a_{1}g_{1}^{l^t}+g_2v_{2}^{l^t}+\cdots+a_ng_{n}^{l^t}=0.$$
Then
$$a_{1}^{l^{m-t}}g_{1}+a_2^{l^{m-t}}g_{2}+\cdots+a_n^{l^{m-t}}g_{n}=0.$$
Since $g_{1},g_{2}, \ldots, g_{n}$ are linearly independent over $\mathbb{F}_{l}$, $a_{1}^{l^{m-t}}=a_2^{l^{m-t}}=\cdots=a_n^{l^{m-t}}=0$. Hence  $a_{1}=a_2=\cdots=a_n=0$. This is a contradiction. Thus,  $g_{1}^{l^t},g_{2}^{l^t}, \ldots, g_{n}^{l^t}$ are linearly independent over $\mathbb{F}_{l}$.

Next, by Theorem \ref{th:Gab2} and Corollary \ref{co:Gab}, $C$ is an MRD code with parameters $[n,k,n-k+1]$, and $C$ is also an MDS code with parameters $[n,k,n-k+1]$ over $\mathbb{F}_{q}$.
\qed

\begin{thm} \label{th:4.11} Let $n\leq m$. If $0\leq t\leq k_1-1$ and $k_1-t+1\leq k_2\leq m-t$, then

$(1)$ there exists an EAQEC code with parameters $[[n,t,\mathrm{min}\{n-k_1+1,k_2+1\};k_2-k_1+t]]_{q}$.

$(2)$ when $n-k_1=k_2$, there exists an EAQEC MDS code with parameters $[[n,t,n-k_1+1;k_2-k_1+t]]_{q}$.
\end{thm}
\pf Taking $$G_1=M_{k_1}(g_1,g_2,\ldots,g_n)=\begin{pmatrix}g_1 & g_2 & \ldots &g_n \\
     g_1^{[1]} & g_2^{[1]} & \ldots &g_n^{[1]}\\
     g_1^{[2]} & g_2^{[2]} & \ldots &g_n^{[2]}\\
    \vdots& \vdots & \ddots & \vdots \\
   g_{1}^{[(k_1-1)]} & g_2^{[(k_1-1)]} & \ldots &g_n^{[(k_1-1)]}\\
  \end{pmatrix}.~~~~~~~~~~~~~~~~~~~~~~~$$
Let $C_1$ be a linear code with the generator matrix $G_1$. Then, by Theorem \ref{th:Gab2} and Corollary \ref{co:Gab}, $C_1$ is an MDS code with parameters $[n,k_1,n-k_1+1]$ over $\mathbb{F}_{q}$.

Let $\tilde{g_1}=g_1^{l^t},\tilde{g_2}=g_2^{l^t},\ldots,\tilde{g_n}=g_n^{l^t}$.
Set, $$H_2=M_{k_2}(\tilde{g_1},\tilde{g_2},\ldots,\tilde{g_n})=\begin{pmatrix}\tilde{g_1}& \tilde{g_2}& \ldots &\tilde{g_n} \\
     \tilde{g_1}^{[1]} & \tilde{g_2}^{[1]} & \ldots &\tilde{g_n}^{[1]}\\
     \tilde{g_1}^{[2]} & \tilde{g_2}^{[2]} & \ldots &\tilde{g_n}^{[2]}\\
    \vdots& \vdots & \ddots & \vdots \\
   \tilde{g_{1}}^{[(k_2-1)]} & \tilde{g_2}^{[(k_2-1)]} & \ldots &\tilde{g_n}^{[(k_2-1)]}\\
  \end{pmatrix}.~~~~~~~~~~~~~~~~~~~~~~~$$
Suppose that $C_2$ is a linear code with the parity-check matrix
$H_2$ . Then,  by Corollary \ref{co:Gab1}, $C_2$ is an MDS code with parameters $[n,n-k_2,k_2+1]$ over $\mathbb{F}_{q}$.

Since $0\leq t\leq k_1-1$ and $k_1-t+1\leq k_2\leq m-t$, we have
$$c=\mathrm{rank}\begin{pmatrix}G_1\\H_2^{(p^{e-0})}\end{pmatrix}-k_1=k_2-k_1+t.$$
Thus, by Theorem \ref{th:4.1}, there exists an EAQEC code with parameters $[[n,t,\mathrm{min}\{n-k_1+1,k_2+1\};k_2-k_1+t]]_{q}$.

$(2)$ When $n-k_1=k_2$, according to $(1)$,  there is an EAQEC code with parameters $[[n,t,d;k_2-k_1+t]]_q$, where $d=\mathrm{min}\{n-k_1+1,k_2+1\}=n-k_1+1$.

Since $2(d-1)=2(n-k_1)=n-t+(k_2-k_1+t)$,  there is an EAQEC MDS code with parameters  $[[n,t,n-k+1;k_2-k_1+t]]_{q}$.
\qed

\begin{exam} By Theorem \ref{th:4.11}, taking some special $q$, we obtain  new
EAQEC MDS codes  in Table 2. Compared to the EAQEC MDS codes in \cite{Liu1}, we have that the number $c$ of entanglement bits of our EAQEC MDS codes  obtained in Table $2$ are smaller than all of them.
\begin{table}
\caption{MDS EAQEC codes comparison.}
\begin{center}\begin{tabular}{|c|c|c|}
\hline
$q$&New  EAQEC MDS codes & EAQEC MDS codes from Corollary $3.19$ \cite{Liu1} \\
\hline
$11^5$&$[[5,2,3;1]]_{11^5}$&$[[5,2,4;3]]_{11^5}$\\
$13^6$&$[[6,2,4;2]]_{13^6}$&$[[6,2,5;5]]_{13^6}$\\
$17^8$&$[[8,4,4;2]]_{17^8}$&$[[8,4,5;4]]_{17^8}$\\
\hline
\end{tabular}\end{center}
\end{table}
\end{exam}

\section{Code comparisons and conclusions}
In this paper, we have developed a new method for  constructing EAQEC codes by using generator matrix of one code and parity check matrix of another code over finite field $\mathbb{F}_{q}$. Using this method, we have constructed three clasess of new EAQEC MDS codes. In Table 3, we give our general conclusions to make comparisons with those known results in Refs.\cite{Chen1,FanJ,Lu,Liu1,Mu,Qian,Qian2,Kor,Wang}. The results show that the lengths and entanglement bits of those known conclusions above EAQEC MDS codes studied in the literatures are  fixed.  However,  the lengths of two classes of EAQEC MDS codes derived from our construction are very flexible, and the entanglement bits of three classes of EAQEC MDS codes derived from our construction are very flexible.

\begin{sidewaystable}
   \caption{comparisons of EAQEC MDS codes.}
\scriptsize
   \begin{tabular}{llll}
   \hline
    Parameters $[[n,k,d;c]]_q$ & Constraints & Distance & References\\
   \hline
   $[[\frac{q-1}{at},\frac{q-1}{at}-2d+6,d;4]]_q$ & $q=l^2$, $l=atm+1$ is an odd prime power,\\& $a$ be even, or $a$ be
odd and $t$ be even & $\frac{at}{2}m+2\leq d\leq(\frac{at}{2}+3)m+1 $ &\cite{Lu}\\
    $[[\frac{q+1}{10},\frac{q+1}{10}-2d+7,d;5]]_q$ & $q=l^2, l\equiv 7~(\mathrm{mod}~10)$, $1\leq \lambda \leq\frac{l+3}{10}$& $d=\frac{3}{5}(l-7)+2\lambda+4 $ &\cite{Kor}\\
   $[[\frac{q+1}{5},\frac{q+1}{5}-2d+6,d;4]]_q$ & $q=l^2,l=2^a, l\equiv 2~(\mathrm{mod}~10)$, $1\leq \lambda \leq\frac{l+3}{5}$& $d=\frac{3}{5}(l-2)+2\lambda+1 $ &\cite{Kor}\\
   $[[q+1,q+1-2d+6,d;4]]_q$ &$q=p^{2a},p^a\equiv1~(\mathrm{mod}~4)$ &$p^a+3\leq d \leq3p^a-1$ and $d$ is even & \cite{Chen1}\\
   $[[q+1,q+1-2d+3,d;1]]_q$ & $q=p^{2a}$&$2\leq d \leq p^a$ and $d$ is even & \cite{FanJ}\\
   $[[q+1,q-2d+11;d,9]]_q$& $q=p^{2a}, p^a\equiv3~(\mathrm{mod}~4), p^a>7$ &$2p^a+4\leq d\leq4p^a-2$ even &\cite{Mu}\\
    $[[q+1,q-d+2,d;d-1]]_q$ &$q=p^{2a}, r\mid p^a-1$ and $r\nmid p^a+1$& $2\leq d \leq\frac{(r-1)(p^{2a}-1)}{2}+2$&\cite{Qian}\\
     $[[q+1,q-2d+4\alpha(\alpha-1)+3,d;1+4\alpha(\alpha-1)]]_q$ & $q=p^{2a}$, $p$ is odd ,$\alpha\in[1,\frac{p^a+1}{4}]$ &  $2+2(\alpha-1)(p^a+1)\leq d \leq2+2\alpha(p^a-1)$ even &\cite{Wang}\\
    $[[q+1,q-2d+4\alpha^2+2,d;4\alpha^2]]_q$ & $q=p^{2a}$, $p$ is odd,$\alpha\in[1,\frac{p^a-1}{4}]$& $2+(2\alpha-1)(p^a+1)\leq d \leq2+(2\alpha+1)(p^a-1)$  even &\cite{Wang}\\
    $[[q+1,q-2d+4\alpha^2+2,d;4\alpha^2]]_q$ &$q=p^{2a}$,$p$ is even,$\alpha\in[1,\frac{p^a-1}{4}]$& $2+(2\alpha-1)(p^a+1)\leq d \leq2+(2\alpha+1)(p^a-1)$  odd &\cite{Wang}\\
    $[[q+1, q-2d+4\alpha(\alpha-1)+3,d;1+4\alpha(\alpha-1)]]_q$ &$q=p^{2a}$,$p$ is even ,$\alpha\in[1,\frac{p^a+1}{4}]$& $2+2(\alpha-1)(p^a+1)\leq d \leq2+2\alpha(p^a-1)$ even &\cite{Wang}\\
    %$[[n,n-d+1,d;d-l]]_q$& $\mathrm{gcd}(p^{2a-k}+1,p^{2a}-1)=s$,\\&$p^{2a}-1=st,n\leq p^{2a}+1$&$n-\mathrm{min}\{n,t\}+1<d<n+1$&\cite{Liu1}\\
    % $[[n,n-d+1,d;d-l]]_q$& $\mathrm{gcd}(p^{2a-k}+1,p^{2a}-1)=s$,\\&$p^{2a}-1=st,n\leq p^{2a}+1$&$1\leq d<\mathrm{min}\{n,t\}+1$&\cite{Liu1}\\
     %$[[2l,l,d;l]]_q$& $p$ is odd, $l\mid p^{2a}-1$,\\&$l\nmid1+p^{k},2(p^k+1)\mid p^{2a}-1$&$d=l+1,l\leq\frac{p^{e}-1}{2}$&\cite{Liu1}\\
     %$[[2l+1,l,d;l+1]]_q$& $p$ is odd, $l\mid p^{2a}-1$,\\&$l\nmid1+p^{k},2(p^k+1)\mid p^{2a}-1$&$d=l+2,l\leq\frac{p^{e}-1}{2}$&\cite{Liu1}\\
     %$[[2l+2,l,d;l+2]]_q$& $p$ is odd, $l\mid p^{2a}-1$,\\&$l\nmid1+p^{k},2(p^k+1)\mid p^{2a}-1$&$d=l+3,l\leq\frac{p^{e}-1}{2}$&\cite{Liu1}\\
     %$[[n,n-d+1,d;d-l]]_q$& $p$ is odd, $r\mid p^{2a}-1$,\\&$r|1+p^{2a-k},n=\frac{p^{2a}-1}{r}$&$2\leq d<n$&\cite{Liu1}\\
     $[[\frac{p-1}{2},\frac{p-1}{2}-2l,d;2l]]_q$&$q=p^e,p\equiv1~(\mathrm{mod} ~4)$& $d=2l+1, 1\leq l\leq\frac{p-5}{4}$& \cite{Liu1}\\
     $[[\frac{p-1}{2},\frac{p-1}{2}-2l+1,d;2l-1]]_q$&$q=p^e, p\equiv3~(\mathrm{mod} ~4)$&$d=2l, 1\leq l\leq \frac{p-3}{4}$& \cite{Liu1}\\
     $[[q+1,q-2d+4m(m-2),d;4(m-1)^2+1]]_q$&$q=l^2$ is an odd prime power, $2\leq m\leq \frac{l-1}{2}$&$d=2(m-1)l + 2$& \cite{Qian2}\\
     $[[q+1,q-2d+4m(m-2),d;4(m-1)^2+1]]_q$&$q=l^2$ and $l=2^a$, \\& $a\geq2$ $2\leq m\leq \frac{l}{2}$&$d=2(m-1)l + 2$& \cite{Qian2}\\
     $[[n,t-1,n-k+1;j-k+t]]_q$&$n\leq q-1, 1\leq t\leq k+1$,\\&$k+1\leq t+j\leq n $,$n-k=1+j$&$d=n-k+1$& Theorem \ref{th:4.3}\\
    $[[q+1,1,q-k+2;q-2k+2]]_q$&$1\leq k<\lceil\frac{q+1}{2}\rceil$&$d=q-k+2$& Theorem \ref{th:4.6}\\
     $[[n,t,n-k_1+1;k_2-k_1+t]]_{q^m}$&$n\leq m, 0\leq t\leq k_1-1$,\\&$k_1-t+1\leq k_2\leq m-t$,$n-k_1=k_2$&$d=n-k_1+1$& Theorem \ref{th:4.11}\\
 \hline
\end{tabular}
\end{sidewaystable}

\textbf{Acknowledgements}
This work was supported by Scientific Research
Foundation of Hubei Provincial Education Department of China. (Grant
No. Q20174503) and the National Science Foundation of Hubei
Polytechnic University of China (Grant No. 17xjz03A).

\end{document}